\documentclass[aps,prl,twocolumn,superscriptaddress,longbibliography]{revtex4-1}

\usepackage{mathrsfs}
\usepackage{tabularx}
\usepackage{slashed}
\usepackage{amsmath}
\usepackage{amsfonts}
\usepackage{graphicx,color}
\usepackage{times}
\usepackage[caption=false]{subfig}
\usepackage[colorlinks,linkcolor=red,citecolor=blue]{hyperref}
\newcommand{\comment}[1]{}
\AtBeginDocument{%
    \newwrite\bibnotes
    \def\bibnotesext{Notes.bib}
    \immediate\openout\bibnotes=\jobname\bibnotesext
    \immediate\write\bibnotes{@CONTROL{REVTEX41Control}}
    \immediate\write\bibnotes{@CONTROL{%
    apsrev41Control,author="08",editor="1",pages="1",title="0",year="1"}}
     \if@filesw
     \immediate\write\@auxout{\string\citation{apsrev41Control}}%
    \fi
}%

\begin{document}

\title{Broken Symmetry in Ideal Chern Bands}

\author{Hui Liu}
\affiliation{Department of Physics, Stockholm University, AlbaNova University Center, 106 91 Stockholm, Sweden}

\author{Kang Yang}
\affiliation{Dahlem Center for Complex Quantum Systems and Fachbereich Physik, Freie Universit\"at Berlin, 14195 Berlin, Germany}

\author{Ahmed Abouelkomsan}
\affiliation{Department of Physics, Massachusetts Institute of Technology, Cambridge, Massachusetts 02139, USA}

\author{Zhao Liu}\thanks{zhaol@zju.edu.cn}
\affiliation{Zhejiang Institute of Modern Physics, Zhejiang University, Hangzhou 310058, China}

\author{Emil J. Bergholtz}\thanks{emil.bergholtz@fysik.su.se}
\affiliation{Department of Physics, Stockholm University, AlbaNova University Center, 106 91 Stockholm, Sweden}

\date{\today}
\begin{abstract}
  Recent observations of the fractional anomalous quantum Hall effect in moir\'e materials have reignited the interest in fractional Chern insulators (FCIs). The chiral limit in which analytic Landau level-like single-particle states form an ``ideal" Chern band and local interactions lead to Laughlin-like FCIs at $1/3$ filling, has been very useful for understanding these systems by relating them to the lowest Landau level. We show, however, that, even in the idealized chiral limit, a fluctuating quantum geometry is associated with strongly broken symmetries and a phenomenology very different from that of Landau levels. In particular, particle-hole symmetry is strongly violated and e.g. at $2/3$ filling an emergent interaction driven Fermi liquid state with no Landau level counterpart is energetically favoured. 
  In fact, even the exact Laughlin-like zero modes at $1/3$ filling have a non-uniform density tracking the underlying quantum geometry. Switching to a Coulomb interaction, the ideal Chern band with electron filling of $1/4$ features trivial charge density wave states. Moreover, applying a particle-hole transformation reveals that the ideal Chern band with hole filling of $3/4$ supports a quantum anomalous Hall crystal with quantized Hall conductance of $e^2/h$. These phenomena have no direct lowest Landau level counterpart.
\end{abstract}

\maketitle

\emph{Introduction.} --- The rapid development of techniques to fabricate van-der-Waals heterostructures with moiré patterns is providing excitingly new opportunities to realize and manipulate novel quantum phases of matter~\cite{Geim2013,Novoselov2016}. Fractional Chern insulators (FCIs)~\cite{kolread,tang_high-temperature_2011,PhysRevLett.106.236803,PhysRevLett.106.236804,sheng2011fractional,PhysRevX.1.021014,mollercooper,PhysRevLett.105.215303,bergholtz2013topological,PARAMESWARAN2013816,LIU2024515}, which are lattice analogues of the celebrated fractional quantum Hall (FQH) states in continuum two-dimensional electron gases (2DEGs)~\cite{FQHE,Laughlin,RevModPhys.89.025005}, have been theoretically predicted and experimentally observed in graphene~\cite{PhysRevLett.124.106803,repellinChernBandsTwisted2020,PhysRevResearch.2.023237,zhaoTDBG,xie2021fractional,lu2023fractional} and transition metal dichalcogenide moir\'e materials~\cite{PhysRevResearch.3.L032070,PhysRevB.107.L201109,FCI_MoTe2_1,FCI_MoTe2_2,FCI_MoTe2_3, PhysRevX.13.031037}. The exciting experimental evidence of FCIs has further attracted tremendous theoretical attention to understand the phenomena emerging in experiments~\cite{dong2023theory,zhou2023fractional,PhysRevLett.132.036501,PhysRevB.108.085117,PhysRevResearch.5.L032022,yu2023fractional,abouelkomsan2023band,xu2024maximally,dong_anomalous_2023}.  

Based on the analogy between FCIs and the FQH states, the emergence of FCI states in moir\'e materials suggests the similarity of the partially filled moir\'e band to a Landau level (LL) in 2DEGs \cite{PhysRevResearch.2.023237,PhysRevLett.127.246403,morales2023magic}. This similarity imposes strong conditions on the dispersion, topology, and quantum geometry~\cite{Goerbig2012,PhysRevB.90.165139,jackson2015geometric,PhysRevLett.114.236802,PhysRevB.104.045103,PhysRevB.104.045104,SciPostPhys.12.4.118,PhysRevB.104.115160,PhysRevLett.127.246403,PhysRevResearch.2.023237,PhysRevLett.128.176403,PhysRevLett.128.176404,10.21468/SciPostPhys.12.1.018,PhysRevResearch.5.L012015,PhysRevResearch.5.L032048,Vortexability_band,PhysRevA.108.032218} of the moir\'e band. In general, a moir\'e band is expected to mimic a LL if (i) it is nearly flat in dispersion; (ii) it carries a nonzero Chern number; (iii) its quantum geometry, characterized by the Berry curvature and the Fubini-Study (FS) metric \cite{Provost1980,Ran2013}, is nearly uniform in the entire Brillouin zone (BZ); and (iv) its FS metric varies approximately in sync with the Berry curvature. The last two conditions originates from the special quantum geometry of a LL, i.e., the Berry curvature $\Omega$ and the FS metric $g$ are constant and related to each other by $g_{ij}=\left(n+\frac{1}{2}\right)|\Omega|\delta_{ij}$, where $n$ is the LL index~\cite{PhysRevB.104.045103}.

\begin{figure}
\centering
\includegraphics[width=\linewidth]{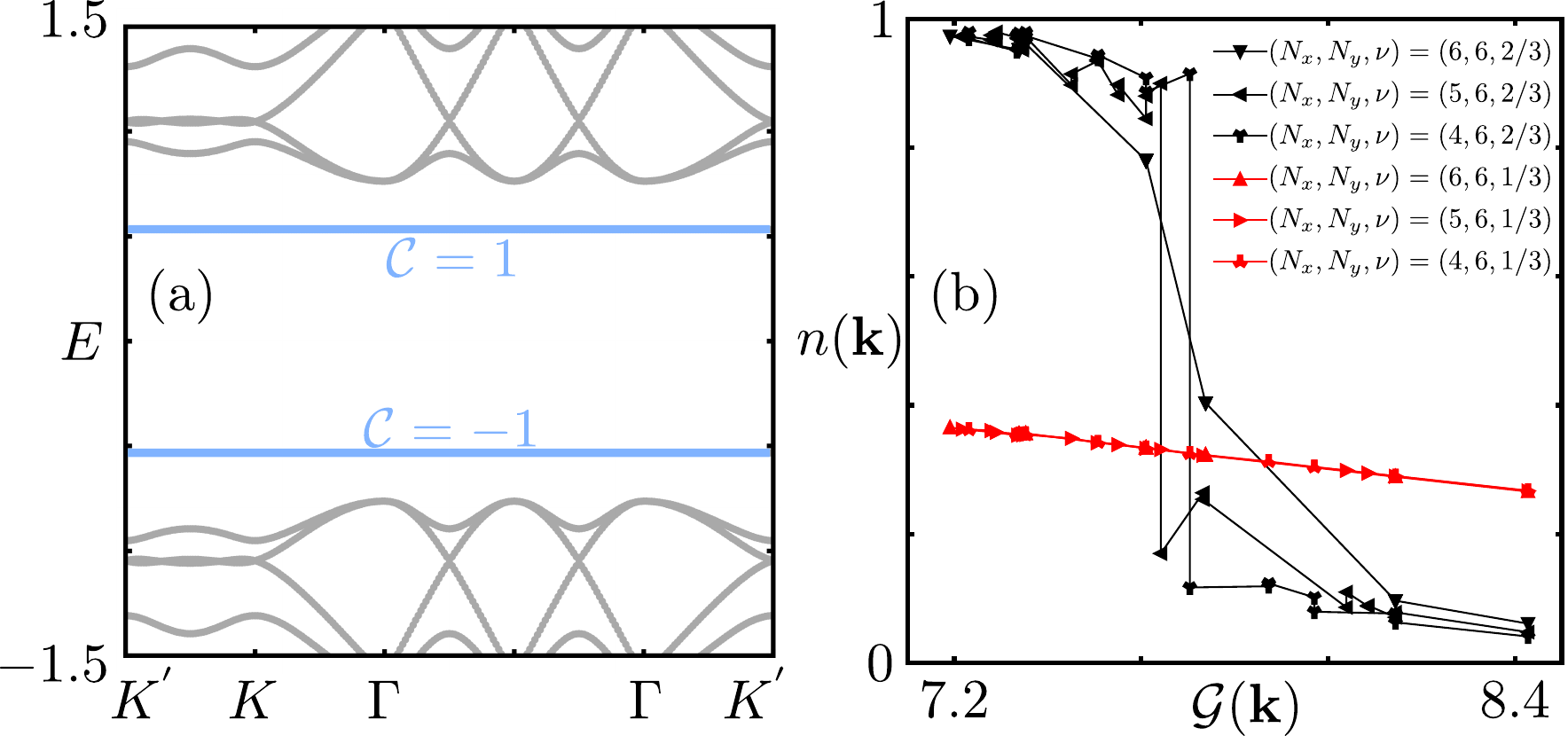}
\caption{{\bf Broken symmetry in an ideal Chern band}. (a) Moir\'e band structure of cTBG at the first magic angle. The horizontal lines are the two flat bands with Chern number $\mathcal C=\pm 1$. (b) Our main result: the occupation number of the many-body ground state $n(\mathbf{k})$ as a function of the quantum geometry, $\mathcal{G}(\mathbf{k})\equiv\text{log}[\text{tr}[ g(\mathbf{k})]]=\text{log}[|\Omega(\mathbf{k})|]$, for a FCI state at filling $\nu=1/3$ of the $\mathcal C =-1$ band and an emergent Fermi liquid at $\nu=2/3$. This shows that ideal Chern bands strongly breaks both the reciprocal-space translation symmetry and the particle-hole symmetry that are present in Landau levels. 
\label{fig:main_result}
}
\end{figure}

According to the criteria above, twisted bilayer graphene in the chiral limit (cTBG)~\cite{PhysRevLett.122.106405} stands out as an extremely promising platform to realize the LL physics in moir\'e systems. The valence and conduction bands of cTBG are perfectly flat at magic twist angles~\cite{PhysRevLett.122.106405} and carry unit Chern numbers. Furthermore, their quantum geometry are ``ideal''~\cite{PhysRevResearch.2.023237,PhysRevLett.127.246403} in the sense that the FS metric varies exactly in sync with the Berry curvature at each ${\bf k}$ point of the BZ via $g_{ij}({\bf k})=\frac{1}{2}|\Omega({\bf k})|\delta_{ij}$, although $g({\bf k})$ and $\Omega({\bf k})$ still fluctuate with ${\bf k}$~\cite{PhysRevResearch.2.023237}. The cTBG bands are hence almost identical to the $n=0$ lowest LL (LLL) except that their quantum geometry is not uniform. These elegant properties lead to LLL-like single-body wave functions, and exact zero modes for pseudopotential-like short-range interactions in cTBG flat bands, just like the Laughlin states in the LLL~\cite{PhysRevLett.127.246403}. The ideal quantum geometry also emerges for flat bands with higher Chern number in the chiral limit of multilayered graphene~\cite{PhysRevLett.128.176403,PhysRevLett.128.176404}. While the aforementioned effective continuum descriptions are most relevant for moir\'e systems, the Kapit-Mueller model~\cite{PhysRevLett.105.215303} and arbitrary Chern number generalizations thereof~\cite{PhysRevLett.116.216802} provide analogous ideal band geometry in tight-binding models and exact parent Hamiltonians for bosonic states. In general, ideal Chen bands have been described as LLLs in curved space \cite{PhysRevResearch.5.L032048}. 
 
Given the glaring similarity between the ideal flat bands in cTBG [cf. Fig.~\ref{fig:main_result}(a)] and the LLL, 
one may naturally expect that they exhibit similar phases at the corresponding fillings. 
However, in this work we establish that cTBG has a much richer phase diagram.
We find that the particle-hole symmetry in a single cTBG flat band is still strikingly broken by the non-uniform, albeit ideal quantum geometry [see Fig.~\ref{fig:main_result}(b)]. Remarkably, even a pseudopotential-like local interaction can induce an unconventional Fermi liquid state at band filling $\nu=2/3$ with no LL counterpart~\cite{PhysRevLett.111.126802}. The ground state at $\nu=1/3$ remains an exact zero-energy Laughlin-like state --- which, however, also exhibits a non-uniform particle density caused by the fluctuating FS metric tensor [Fig.~\ref{fig:main_result}(b)]. 
Furthermore, the Coulomb interaction favors a charge density wave instead of the composite Fermi liquid at electron filling $\nu=1/4$ and a quantum anomalous Hall crystal at hole filling $\nu=3/4$, which again distinguish the cTBG ideal bands from the LLL.

\emph{Setup.} --- 
We focus on the cTBG~\cite{PhysRevLett.122.106405}. At first magic angle, the spin and valley polarized Hamiltonian provides two perfectly flat bands with opposite Chern number $\pm1$ [see Fig.~\ref{fig:main_result}(a)] and possess ideal quantum geometry. 
We consider the interacting physics in the lower flat band. We project the electron-electron interaction to this single valence band and reach
\begin{eqnarray}
H^{\text{proj}}=\sum_{\mathbf{k}}V_{\mathbf{k}_1\mathbf{k}_2\mathbf{k}_3\mathbf{k}_4}c^\dagger_{\mathbf{k}_1}c^\dagger_{\mathbf{k}_2}c_{\mathbf{k}_3}c_{\mathbf{k}_4},
\end{eqnarray}
with $c^\dagger_{\mathbf{k}}$ ($c_{\mathbf{k}}$) being the fermionic creation (annihilation) operator with momentum $\mathbf{k}$ in the valence band. The normal ordering convention here is consistent with obtaining exact zero-modes at $\nu=1/3$ \cite{PhysRevResearch.2.023237}. The band eigenvectors have been 
encoded into the matrix element $V_{\mathbf{k}_1\mathbf{k}_2\mathbf{k}_3\mathbf{k}_4}$ (see the Supplemental Material~\cite{SupMat} for more details). 
We use exact diagonalization to extract the low-lying energy spectrum and the associated eigenstates for finite systems including $N$ electrons in $N_s$ moir\'e unit cells at band filling $\nu=N/N_s$, which can provide direct evidence to distinguish FCIs from competing phases, such as Fermi liquids and charge density waves, etc.

\begin{figure}
\centering
\includegraphics[width=\linewidth]{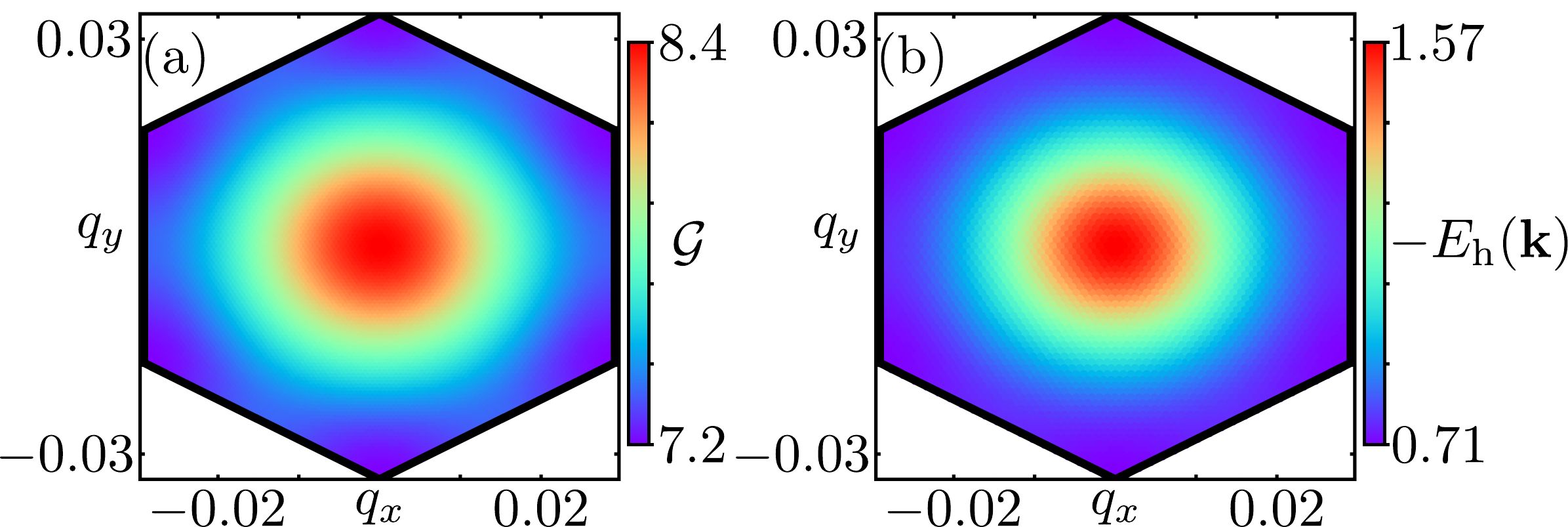}
\caption{{\bf Quantum geometry and single hole dispersion.} (a) The quantum geometry $\mathcal{G}(\mathbf{k})$ and (b) the single hole energy $-E_{\text{h}}(\mathbf{k})$ as a function of momentum $\mathbf{k}=(q_x, q_y)$.
\label{fig:bc_eh}
}
\end{figure}

\emph{Broken symmetry in ideal Chern bands.} --- 
Our main result centers around the striking breaking of reciprocal-space translation symmetry and particle-hole symmetry in the ideal flat band of cTBG, even when we employ a pseudopotential-like short-range repulsive interaction whose Fourier transform is $V(\mathbf{q})=V_0-V_1 a_M^2 |\mathbf{q}|^2$ ($a_M$ the moir\'e lattice constant)~\cite{PhysRevLett.51.605, SupMat, PhysRevResearch.2.023237}. While this interaction gives rise to zero-energy Laughlin-like model FCIs at $\nu=1/3$ filling, the particle occupation number $n(\mathbf{k})$ is not a constant at $1/3$ as for the conventional Laughlin state in the LLL. By contrast, $n(\mathbf{k})$ exhibits a clear linear dependence on the quantum geometry characterized by $\mathcal{G}(\mathbf{k})\equiv\log[\text{tr}[ g(\mathbf{k})]]=\log[|\Omega(\mathbf{k})|]$ [Fig.~\ref{fig:main_result}(b)]. This variation of $n(\mathbf{k})$ is a signature of the breaking of translation symmetry in the reciprocal space, which further suggests a significant center-of-mass dependence of the model Laughlin-like wavefunction within the ideal flat band~\cite{PhysRevResearch.2.023237,PhysRevLett.127.246403}. Notably, as shown in Fig.~\ref{fig:bc_eh} the quantum geometry in terms of $\mathcal G$ [Fig. \ref{fig:bc_eh}(a)], has a similar structure in the moir\'e BZ as the energy of a single hole [Fig. \ref{fig:bc_eh}(b)], thus providing a physical rationale for the inhomogeneous (reciprocal space) occupation density $n(\mathbf k)$ \cite{PhysRevResearch.5.L012015}.  

Moving to filling $\nu=2/3$, we find that the occupation number $n(\mathbf{k})$ completely deviates from the value $2/3$ of the $\nu=2/3$ FQH state in the LLL. Instead, as displayed in Fig.~\ref{fig:main_result}(b), a sharp jump from $n(\mathbf{k})\approx 1$ to $n(\mathbf{k})\approx 0$ signifies the emergence of a Fermi liquid, with a stable Fermi surface around $\mathcal{G}\approx 7.6$. The very distinct behavior of $n(\mathbf{k})$ at $\nu=2/3$ from that at $\nu=1/3$ indicates the breaking of particle-hole symmetry in the ideal cTBG bands. Such particle-hole asymmetry in Chern bands has been identified in various non-ideal models: first in a checkerboard lattice~\cite{PhysRevLett.111.126802} and, recently, in moir\'e systems~\cite{PhysRevLett.124.106803,PhysRevB.108.245159} where it is related to the fluctuating quantum geometry~\cite{PhysRevResearch.5.L012015}. That these effects persist even for electrons with pseudopotential-like interaction in ideal flat bands further emphasizes the fundamental difference between Chern bands and the LLL, and disapproves of the expectation that the LLL physics can be perfectly reproduced in ideal flat bands given ideal quantum geometry and the Laughlin-like wave functions in the latter.

\begin{figure}
\centering
\includegraphics[width=\linewidth]{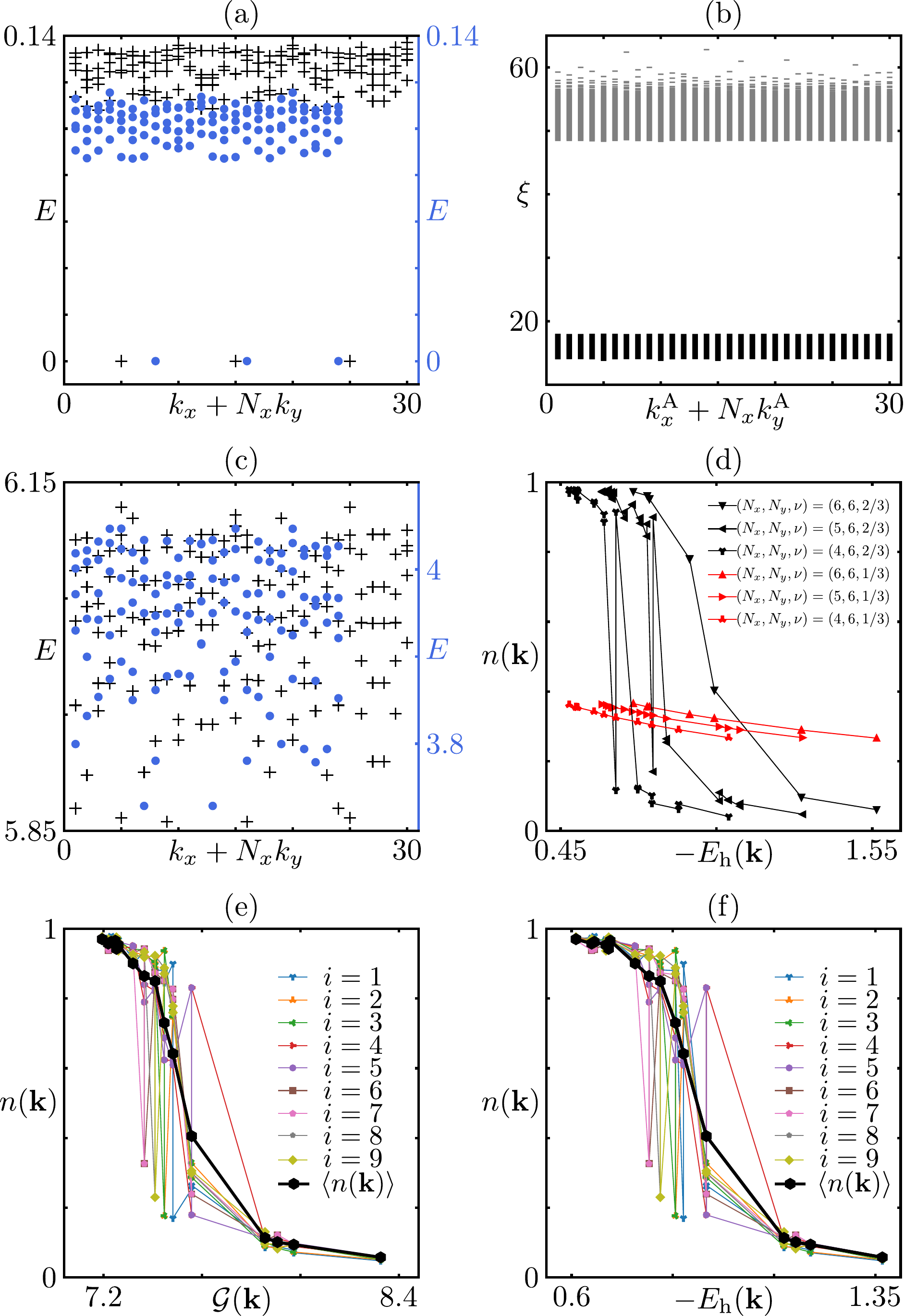}
\caption{{\bf FCIs and Fermi Liquids.} (a) The many-body energy spectrum at band filling $\nu=1/3$ on the $4\times 6$ (blue $\cdot$) and $5\times 6$ clusters (black +), respectively. 
(b) The associated particle-cut entanglement spectrum of the three zero modes on the $5\times 6$ cluster for $N_A=5$. The number of levels in the bottom narrow band is $23256$, which matches the quasihole counting of the $\nu=1/3$ Laughlin state. The very high-energy grey levels are numerical noises. (c) The many-body energy spectrum at band filling $\nu=2/3$. (d) The ground-state occupation number of electrons at $\nu=1/3$ and $\nu=2/3$ as a function of single-hole energy. (e)-(f) The occupation number of electrons for the lowest $9$ excited states at $\nu=2/3$ for the $5\times 6$ cluster as a function of the quantum geometry and the single-hole energy, respectively. Here, the black hexagon and the bold line represent the averaged occupation number over the lowest $20$ energy states.
\label{fig:FCI_broken_symmetry}
}
\end{figure}

\emph{Evidence for the FCIs and the emergent Fermi liquid.} --- We now provide more numerical evidence on the FCIs at $\nu=1/3$ and the Fermi liquid at $\nu=2/3$ in the cTBG ideal band. 
The most straightforward signature of the $\nu=1/3$ Laughlin-like model FCIs is the three exactly zero-energy ground states with momenta predicted by the Haldane statistics~\cite{PhysRevX.1.021014,bernevigEmergentManybodyTranslational2012}. This can be clearly seen in the low-lying energy spectrum for any finite system [Fig.~\ref{fig:FCI_broken_symmetry}(a)]. The ground-state topological order can be further diagnosed by the particle-cut entanglement spectrum~\cite{li_entanglement_2008,sterdyniak_extracting_2011}. 
Here, we divide the whole system into $N_A$ and $N_B=N-N_A$ electrons. 
By tracing out the subsystem $B$, we effectively create holes in the system, so the reduced density matrix of the subsystem $A$, $\rho_A=\text{tr}_B[\frac{1}{3}\sum_{i=1}^{3}|\Psi_i\rangle\langle\Psi_i|]$, contains the information of the quasihole excitations above the ground states $|\Psi_i\rangle$. In Fig.~\ref{fig:FCI_broken_symmetry}(b), we display the particle-cut entanglement spectrum, which is the spectrum of $-\log\rho_A$, for the three zero modes. There is a narrow band at the bottom of the entanglement spectrum, in which the number of levels matches the quasihole counting of the $\nu=1/3$ Laughlin state. More entanglement levels appear at very high entanglement energies above $\xi\approx 50$. However, as these levels correspond to exponentially small eigenvalues of $\rho_A$ beyond the machine precision limit, we attribute them to numerical noise. Therefore, the entanglement spectrum of the zero modes at $\nu=1/3$ in fact has an infinite entanglement gap, which is a remarkable feature of model wavefunctions~\cite{li_entanglement_2008,PhysRevLett.116.216802}. Both the energy spectrum and the ground-state entanglement spectrum strongly indicate the existence of Laughlin-like model FCIs at $\nu=1/3$ and rule out the possibility of a trivial charge density wave.

At band filling $\nu=2/3$, the absence of topological degeneracy in the energy spectrum, as illustrated in Fig.~\ref{fig:FCI_broken_symmetry}(c), already rules out the presence of FCIs. Given the dispersionless nature of the ideal cTBG band, the Fermi liquid feature in Fig.~\ref{fig:main_result}(b) must originate solely from the electron-electron correlation, whose role can be exposed more clearly in the hole picture. It has been pointed out in early works that the electron-electron interaction leads to 
an extra single-hole dispersion $E_{\text{h}}(\mathbf{k})$ after the particle-hole transformation~\cite{PhysRevLett.111.126802,PhysRevLett.124.106803, SupMat}. This hole dispersion in general dominates over the hole-hole interaction at relatively large fillings and gives rise to the correlation-induced Fermi liquid. We confirm the correlated nature of the Fermi liquid in cTBG ideal bands by drawing the ground-state occupation number of electrons as a function of $E_{\text{h}}(\mathbf{k})$. As shown in Fig.~\ref{fig:FCI_broken_symmetry}(d), $n(\mathbf{k})$ at $\nu=2/3$ has an overall trend to drop from $1$ to $0$ following the hole dispersion $E_{\text{h}}(\mathbf{k})$. The zigzags in $n(\mathbf{k})$ for some system sizes result from the hole-hole interaction which distorts the Fermi surface of electrons. 
In contrast, $n(\mathbf{k})$ for the $\nu=1/3$ FCI only weakly varies with the hole dispersion. The Fermi liquid behavior at $\nu=2/3$ even extends beyond the ground state, where excited states possess similar features [see Figs.~\ref{fig:FCI_broken_symmetry}(e) and (f)].

We emphasize that the symmetry breaking in cTBG ideal bands with pseudopotential-like interactions is a phenomenon observed across various fillings. For the filling $\nu\in[1/3, 1/2)$, 
and their complements, our results show a consistent behavior, where an FCI emerges at band filling $\nu$ but disappears at band filling $1-\nu$ (see the SM for more details~\cite{SupMat}, also reference~\cite{PhysRevB.90.245401} therein). 

\emph{Charge density wave and quantum anomalous Hall crystal.} --- 
In previous sections, the striking differences between the ideal cTBG band and the LLL have been revealed at fillings where FQH states are expected in the LLL. Now we consider other fillings where there are no incompressible FQH states in the LLL. As a concrete example, we consider $\nu=1/4$, where the Coulomb interaction $V_{\text{C}}(\mathbf{q})\sim 1/|\mathbf{q}|$ in the LLL were found to stabilize the Fermi sea of composite fermions~\cite{ShayeganQuarterCF,WangQuarterCF} (except in the wide quantum well which is beyond the scope of our work). We suppose the electrons in the cTBG interact via the bare Coulomb potential also. Surprisingly, at $\nu=1/4$ of the ideal cTBG band, we observe a charge density wave (CDW) rather than the Fermi sea of composite fermions. 

\begin{figure}
\centering
\includegraphics[width=\linewidth]{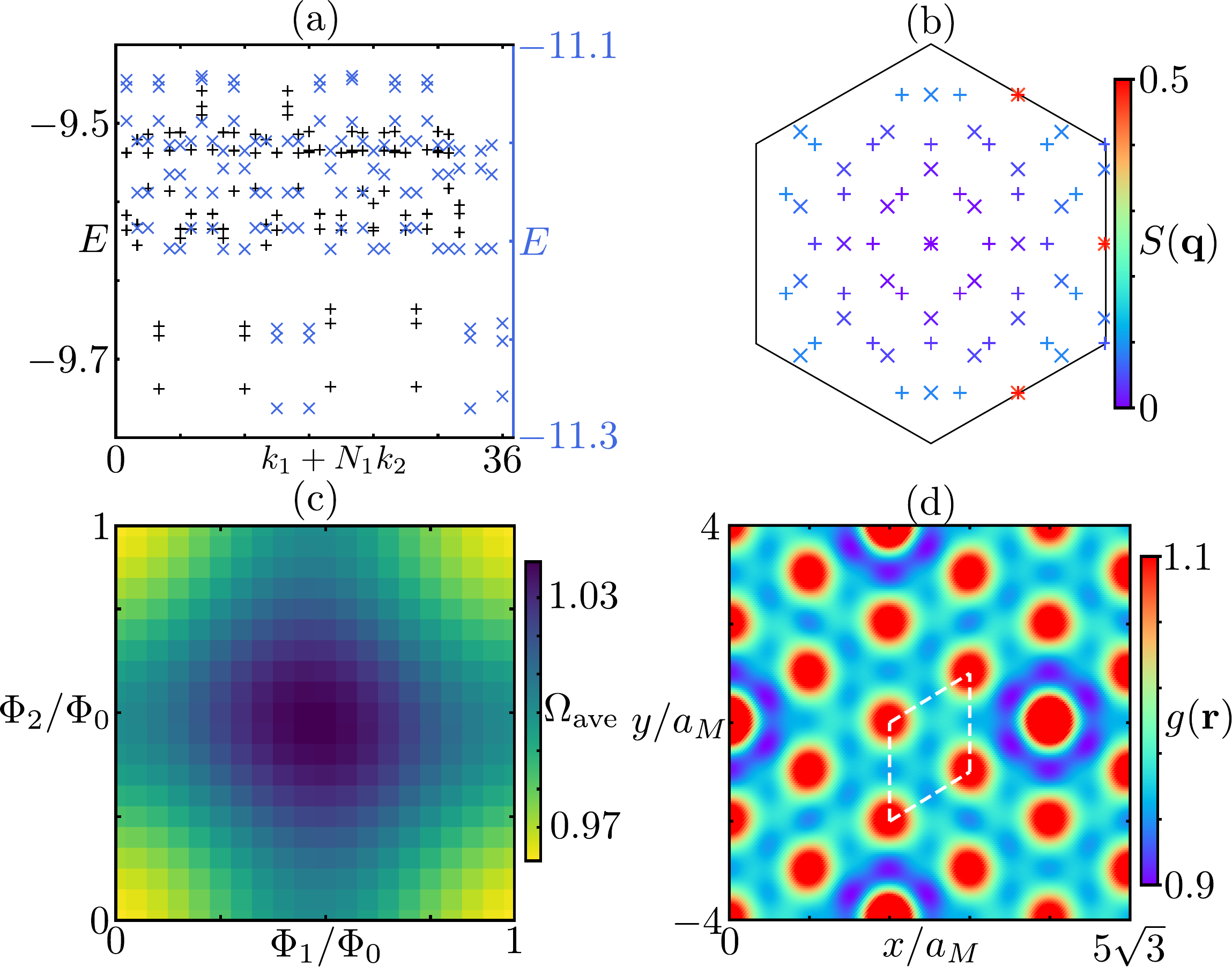}
\caption{{\bf CDWs and quantum anomalous Hall crystals from the Coulomb interaction in the ideal cTBG band}. (a) The many-body energy spectrum at band filling $\nu=1/4$ for the bare Coulomb interaction. We calculate three lowest levels in each momentum sector.  
(b) The ground-state structure factor in the MBZ. 
Here, we use tilted samples to include the three inequivalent ${\bf M}$ points. The $+$ (black in (a)) and $\times$ (blue in (a)) symbols correspond to clusters with $32$ and $36$ moir\'e unit cells, respectively~\cite{SupMat}. 
(c) The many-body Berry curvature distribution of the anomalous Hall crystal phase in clusters with $32$ moir\'e unit cells at band filling $\nu=3/4$ for holes. And (d) the corresponding pair correlation function of ground states. Here, the white dashed rhombus indicates $2a_M\times 2a_M$ CDW unit cells. Here, the averaged Berry curvature is renormalized regarding the $20\times 20$ flux grids to show the stability of the ground states and the quantization of the many-body Chern number.
\label{fig:cdw}
}
\end{figure}

We verify the CDW phase from two aspects. For finite samples whose single-electron momenta allowed by the periodic boundary conditions include the three inequivalent ${\bf M}$ points of the MBZ, we find four approximately degenerate ground states in the energy spectrum separated by the momenta ${\bf M}_{i=1,2,3}$ [Fig.~\ref{fig:cdw}(a)]. This is a strong signal of the charge distribution with order momenta ${\bf M}_i$. The charge order can be diagnosed more reliably by the structure factor
$S(\mathbf{q})=\frac{1}{N_s}(\langle \rho (\mathbf{q})\rho (-\mathbf{q})\rangle-N^2\delta_{{\bf q},{\bf 0}})$,
where $\rho_{\bf q}$ is the density operator projected to the ideal cTBG band. Remarkably, we find pronounced peaks only at the three ${\bf M}$ points [Fig.~\ref{fig:cdw}(b)], confirming the ground states as ${\bf M}$-CDWs, whose unit cell is quadrupled moir\'e unit cell~\cite{PhysRevB.103.125406}. 

Building on recent progress in anomalous Hall crystals— states that combine fractionalized topological order with broken translational symmetry~\cite{polshyn2022topological, hall_crystal1, hall_crystal2, hall_crystal3, hall_crystal4,hall_crystal5,PhysRevLett.133.206504,PhysRevLett.133.206502}—we suggest that such a novel phase could, in principle, exist in an ideal Chern band. At the electron filling $\nu=1/4$, the four ground states are topologically trivial, characterized by an averaged many-body Chern number $\mathcal{C}_{\text{avg}}=0$. 
However, as shown in Fig.~\ref{fig:cdw}(c), the system at hole filling $\nu=3/4$ now has a quantized conductance of $\mathcal{C}_{\text{avg}}=1$ [see SM~\cite{SupMat}, also references~\cite{PhysRevB.55.1142, RevModPhys.88.035005} therein]. Consequently, the pair correlation function shows a $2a_M\times 2a_M$ unit cell structure in Fig.~\ref{fig:cdw}(d), confirming the emergence of an anomalous Hall crystal phase. Such a behavior is reminiscent of reentrant integer quantum Hall effects usually taking place at higher LLs \cite{PhysRevLett.88.076801,PhysRevLett.89.136804,PhysRevLett.122.026802}, which requires disorder to be stabilized in the LLL \cite{PhysRevLett.105.076803}. We note that Hall crystals theorized in the context of LLs spontaneously break continuous translation symmetry leading to gapless modes \cite{PhysRevB.39.8525} in contrast to the cTBG setting where the discrete moiré lattice symmetry is broken leading to gapped anamalous Hall crystals \cite{hall_crystal1}.

Our results demonstrate that both trivial and nontrivial CDWs can survive even in the ideal cTBG band. The fluctuation of the Berry curvature and the quantum metric tensor of the ideal cTBG band, which distinguishes it from the LLL, could be a major mechanism for the formation of these CDWs that are absent in the LLL. 

\emph{Discussion.} --- By employing both a pseudopotential-like interaction and a Coulomb interaction, we have discovered that, even the topological flat band in cTBG exhibits an ideal quantum geometry, yet significant dissimilarities persist when compared to the LLL. 
In the former scenario, the breaking of particle-hole symmetry at band filling $\nu=2/3$ leads to a correlation-induced Fermi liquid, while at $\nu=1/3$ a FCI is supported although also this state has a non-uniform occupation density adapting to the underlying geometry. The breakdown of correspondence between the flat band in cTBG and the LLL also provides hints to other possible competing orders in this ideal setting. In particular, we unveiled both a trivial CDW at electron band filling $\nu=1/4$ and a topological crystal at (normal ordered) hole band filling $\nu=3/4$ in the latter case, characterized by three clear peaks in the structure factor. This contrasts with the anticipated gapless composite Fermi liquid of the LLL physics, highlighting compelling deviations in cTBG. In conclusion, while the ideal cTBG band does indeed elegantly preserve the presence of zero modes of certain short range interactions, the phase diagram is qualitatively altered compared to that of the LLL. 

It should be noted that the quantum geometric quantities are of limited fundamental importance in lattice models due to their dependence on the embedding of the orbitals into real space. 
A more orbital-independent approach should go back to continuum models obtained from first principle calculations. In moir\'e systems, the underlying (e.g., graphene) orbitals are highly localized compared to the moir\' e length scale. Thus the system has a natural continuum description that for practical purposes alleviates the fundamental ambiguity of lattice embeddings.

We also note that normal ordering prescription matters. We have mainly used a convention used in the literature, normal ordering with respect to an empty vacuum, which leads to Laughlin-like zero modes of electrons at $\nu=1/3$ \cite{PhysRevResearch.2.023237}. If instead normal ordered with respect to charge neutrality, the interaction Hamiltonian differs by extra single-particle terms which can be thought as Hartree Fock corrections from the remote bands \cite{PhysRevB.103.205413} but the particle-hole asymmetry within the single band still exists hence our results still hold with the difference that FCIs are no longer exact zero modes of the pseudopotential. 
Remarkably, the alternative normal ordering prescription also leads to anomalous Hall crystals in presence of long range Coulomb interactions even within the ideal band.
We stress that the particle-hole asymmetry considered here is the one emerging within a given (ideal) band which is different from the question of symmetry of the model as a whole. 

The glaring differences between ideal Chern bands and Landau levels identified in this work prompt a natural question that, what criteria can accurately determine to what extent an ideal moir\'e Chern band aligns with the physics of the LLs more generally. 
Understanding this question becomes crucial in realizing non-Abelian topological phases and their associated quantum computing applications in moir\'e systems. It has been experimentally demonstrated that FCIs can be realized without needing a high magnetic field and exhibit robustness at relatively high temperatures. 

While FCIs have been experimentally observed in twisted bilayer graphene \cite{xie2021fractional} at fillings $\nu = 4 - 1/3$ and $\nu = 4 - 2/3$ of the conduction band, we would like to stress that our emphasis here is different as we are studying the spin and valley polarized limit of chiral TBG. It is believed that the state at $\nu = 4 - 1/3$ is spin and valley polarized \cite{parker2021field} while the state at $\nu = 4 - 2/3$ (in the chiral limit) is spin-singlet \cite{repellinChernBandsTwisted2020}. Our results are concerned with particle-hole asymmetry in a band with a single flavor. In the presence of multi-component degrees of freedom, the single-hole dispersion takes a more complicated form than the form in \cite{suppm} therefore the conclusions can differ. It is also possible that external magnetic fields could diminish the existing particle-hole asymmetry at zero magnetic field approaching a more LLL-like limit. We leave this to future study. 

On the other hand, robust spin-valley polarization has been observed in moir\'e TMD systems~\cite{FCI_MoTe2_1,FCI_MoTe2_2,FCI_MoTe2_3, PhysRevX.13.031037} but FCIs were found only at hole doping $\nu_h = 2/3$ of the valence band while absent at the particle-hole dual filling $\nu_h = 1/3$ reflecting strong particle-hole asymmetry similar to our study. While chiral TBG is significantly different from moir\'e TMDs, we believe that our results provide general and relevant insights to this problem. We have shown that even in an idealized model, translation and particle-hole symmetries is strongly violated even in absence of band mixing, hence such symmetries should not be expected in any model derived starting from an underlying lattice structure.

\acknowledgments{{\it Acknowledgements.---} We acknowledge discussions with Jie Wang. H.~Liu and E.~J.~Bergholtz were supported by the Swedish Research Council (VR, grant 2018-00313), the Wallenberg Academy Fellows program of the Knut and Alice Wallenberg Foundation (2018.0460) and the G\"oran Gustafsson Foundation for Research in Natural Sciences and Medicine. Z. ~Liu is supported by the National Natural Science Foundation of China (Grant No. 12350403 and 12374149) and the National Key Research and Development Program of China (Grant No. 2021YFA1401902). K.Yang is supported by the ANR-DFG project (TWISTGRAPH). A. Abouelkomsan is supported by the Knut and Alice Wallenberg Foundation (2022.0348). }

\bibliography{MtcFl}

\pagebreak
\newpage

\onecolumngrid
\pagebreak
\newpage

\section{Supplemental Materials to `Broken Symmetry in Ideal Chern Bands'}

In this supplemental material, we provide the details of the interaction Hamiltonian, the evidence of particle-hole symmetry breaking at other fillings, the spectral flow with respect to magnetic flux insertion, and the choice of samples, and relations between Berry curvature of holes and electrons.

\section{The interaction Hamiltonian and the single-hole dispersion}
\label{sec:interaction_Hamiltonian}

The two-body interaction projected to a single band with specific valley and spin takes the form of 
\begin{eqnarray}
H^{\text{proj}}=\sum_{\mathbf{k}_1,\mathbf{k}_2,\mathbf{k}_3,\mathbf{k}_4}V_{\mathbf{k}_1\mathbf{k}_2\mathbf{k}_3\mathbf{k}_4}c^\dagger_{\mathbf{k}_1}c^\dagger_{\mathbf{k}_2}c_{\mathbf{k}_3}c_{\mathbf{k}_4},
\end{eqnarray}
with the interaction matrix element
\begin{eqnarray}
    V_{\mathbf{k}_1\mathbf{k}_2\mathbf{k}_3\mathbf{k}_4}&=&\frac{1}{2}\sum_{\mathbf{\mathbf{q}}}V(\mathbf{q})\sum_{\tau,\tau'}\sum_{\{m,n\}=-d}^d\delta_{\mathbf{k}_1+\mathbf{k}_2+(m_1+m_2)\mathbf{G}_1+(n_1+n_2)\mathbf{G}_2,\mathbf{k}_3+\mathbf{k}_4+(m_3+m_4)\mathbf{G}_1+(n_3+n_4)\mathbf{G}_2}\cdot\nonumber\\
     &&\delta_{\mathbf{k}_1-\mathbf{k}_4+(m_1-m_4)\mathbf{G}_1+(n_1-n_4)\mathbf{G}_2,\mathbf{q}}\mu^{*}_{m_1,n_1,\tau}(\mathbf{k}_1)\mu^{*}_{m_2,n_2,\tau'}(\mathbf{k}_2)\mu_{m_3,n_3,\tau'}(\mathbf{k}_3)\mu_{m_4,n_4,\tau}(\mathbf{k}_4).
\end{eqnarray}
Here, $V(\mathbf{q})$ indicates either the pseudopotential-like short-range interaction or the bare Coulomb interaction, $\mathbf{G}_1$ and $\mathbf{G}_2$ denote the two moir\'e reciprocal operators, respectively, and $\tau$ ($\tau'$) represents the orbital index. 
$\mu_{m,n}(\mathbf{k})$ is the eigenvector of the valence and topological flat band at $m\mathbf{G}_1+n\mathbf{G}_2$ sector for momentum $\mathbf{k}\in\text{MBZ}$. 
In numerical simulations, we choose a cutoff $d=8$, which is enough to make both the single-particle bandstructure and the many-body energy spectrum converge.

The breaking of particle-hole symmetry is originated from the inhomogeneous single-hole dispersion $E_\text{h}(\mathbf{k})$ (in other words, not a constant). 
Here, we derive the concrete from of $E_\text{h}(\mathbf{k})$. 
The original interaction Hamiltonian in momentum space reads,

By doing the particle-hole transformation $c^\dagger_{\mathbf{k}}\rightarrow c_{\mathbf{k}}$, we have
\begin{eqnarray}
    H^{\text{proj}}\rightarrow \sum_{\mathbf{k}_1,\mathbf{k}_2,\mathbf{k}_3,\mathbf{k}_4}V_{\mathbf{k}_1\mathbf{k}_2\mathbf{k}_3\mathbf{k}_4}c^\dagger_{\mathbf{k}_1}c^\dagger_{\mathbf{k}_2}c_{\mathbf{k}_3}c_{\mathbf{k}_4}+\sum_{\mathbf{k}}E_{\text{h}}(\mathbf{k})c^\dagger_{\mathbf{k}}c_{\mathbf{k}},
\end{eqnarray}
with $E_{\text{h}}(\mathbf{k})=\sum_{\mathbf{k}'}(V_{\mathbf{k}'\mathbf{k}\mathbf{k}'\mathbf{k}}+V_{\mathbf{k}\mathbf{k}'\mathbf{k}\mathbf{k}'}-V_{\mathbf{k}\mathbf{k}'\mathbf{k}'\mathbf{k}}-V_{\mathbf{k}'\mathbf{k}\mathbf{k}\mathbf{k}'})$.

\section{Spectral flow for $\nu=1/3$ and $3/7$} 

\begin{figure*}
\centering
\includegraphics[width=0.5\linewidth]{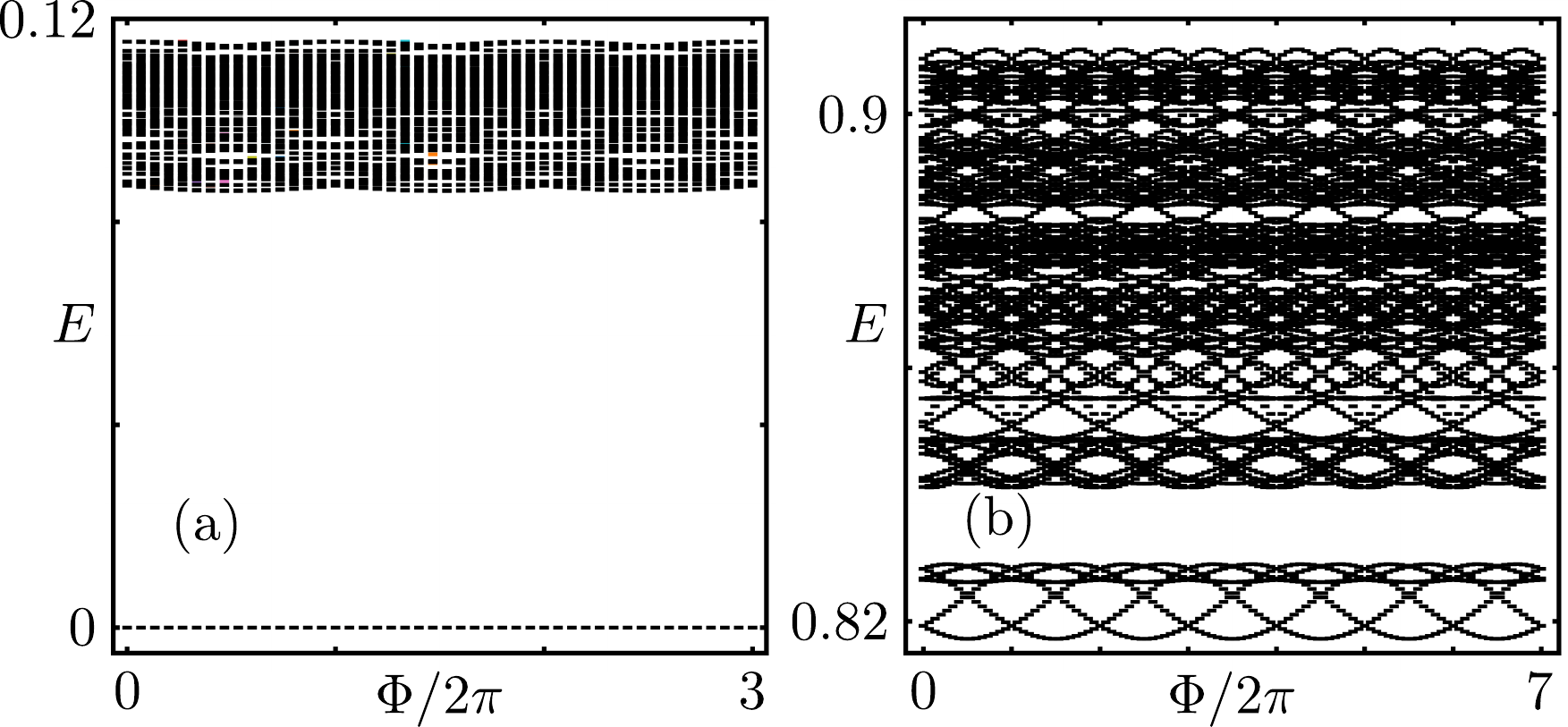}
\caption{The low-lying energy spectrum as a function of magnetic flux $\Phi$ at band filling $v=1/3$ (a) and $3/7$ (b) with a system size $4\times 6$ and $4\times 7$ clusters, respectively. 
Here, we use a pseudopotential-like short-range interaction.
\label{fig:spectral_flow}
}
\end{figure*}

Here, we examine the topological nature of the system at band filling $\nu=1/3$ and $3/7$ by introducing an external magnetic flux. 
As shown in Fig.~\ref{fig:spectral_flow}, for both fillings, the quasi-degenerate FCI ground states remain isolated from the ground states, and evolve into each other with the magnetic flux, which indicates their corresponding fractional transmission.

\section{Results of other fillings}
\label{sec:other_fillings}

\begin{figure*}
\centering
\includegraphics[width=\linewidth]{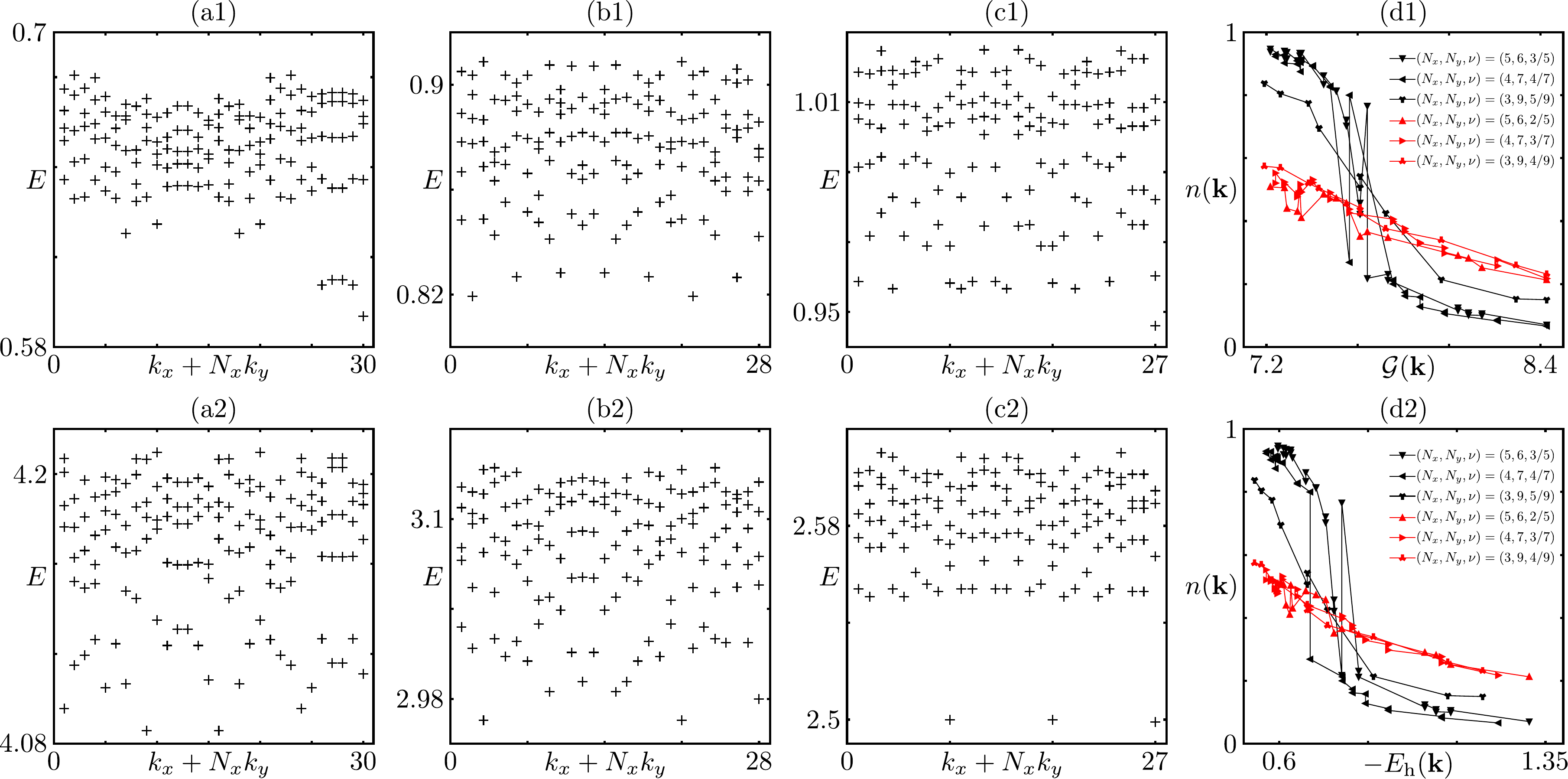}
\caption{(a1-c1) are the low-lying energy spectrum for band filling $\nu=2/5$, $3/7$, and $4/9$, at system size $5\times 6$, $4\times 7$, and $3\times 9$ clusters, respectively. (a2-c2) are the energy spectrum for their complementary band fillings $\nu=3/5$, $3/7$, and $5/9$, respectively. (d1) and (d2) are the occupation number of ground states at different fillings as a function of quantum geometry and the single-hole dispersion, respectively. 
Here, we use a pseudopotential-like short-range interaction and only selective five lowest energies in ED simulations.
\label{fig:other_fillings}
}
\end{figure*}

In this section, we provide a detailed study for other fillings. 
Like the FQHE in the LLL, varying the partial filling of the flat band in cTBG $\nu$, from $1/3$ to $1/2$ (here, we choose $\nu=2/5$, $3/7$, and $4/9$), indeed leads to a energy gap closing between FCI ground states and excitation states, shown in Fig.~\ref{fig:other_fillings}(a1-c1). 

At $\nu=2/5$, five quasi-degenerate ground states appear at momentum sectors satisfying the Haldane statistics and are separated from excitation states with a clear energy gap. 

\begin{figure*}
\centering
\includegraphics[width=0.5\linewidth]{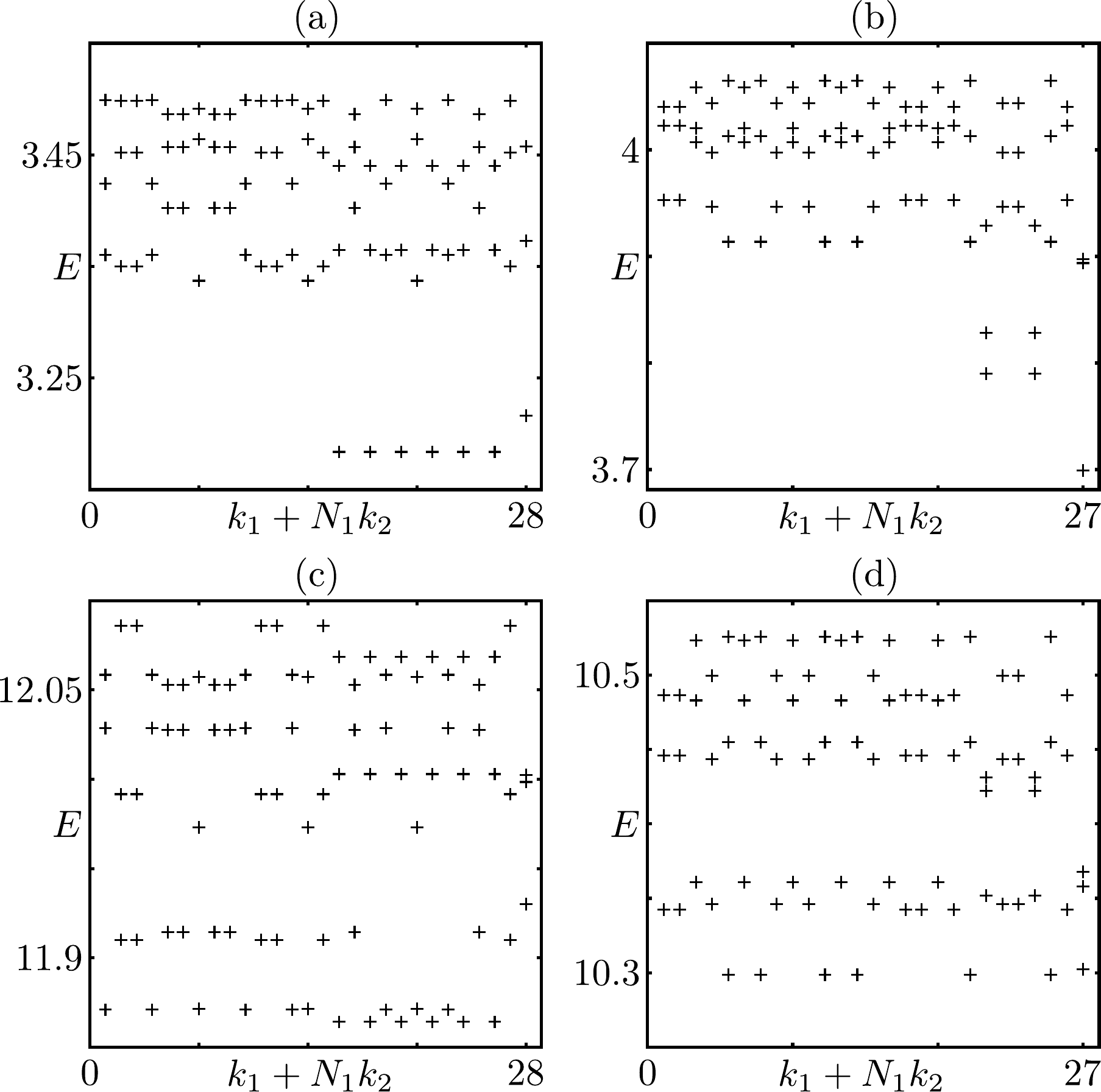}
\caption{(a-b) the low-lying energy spectrum of the tilted sample for banding filling $\nu=3/7$ and $4/9$, at system size $4\times 7$ and $3\times 9$ clusters, respectively. (c-d) are the energy spectrum of their particle-hole partner $\nu=4/7$ and $5/9$, respectively. 
Here, we use a pseudopotential-like short-range interaction and only selective three lowest energies in ED simulations. 
\label{fig:tilted_result}
}
\end{figure*}

For $\nu=3/7$, although the energy gap is less conspicuous, seven quasi-degenerate ground states persist at the right momentum sectors, maintaining isolation from other states. 
This is further validated by introducing a magnetic flux, where the seven ground states never mix with other higher energy states, making the energy gap visible. 
Alternatively, employing a tilted sample configuration reveals the clear separation of the seven ground states from excitation states (see Fig.~\ref{fig:tilted_result})(a).

Approaching $\nu=4/9$, near the phase transition, both clear FCI ground states and energy gap are absent, as evident from both the rectangular sample, shown in Fig.~\ref{fig:other_fillings}(c1) and the tilted sample shown in Fig.~\ref{fig:tilted_result}(c)

We then shift the focus to band filling within $(1/2, 2/3)$, specifically, $\nu=3/5$, $4/7$, and $5/9$, the complimentary partners. 
Surprisingly, the energy spectrum of each of these fillings, both from the rectangular sample and the tilted shows an absence of FCI ground states or an associated energy gap. 
This is again opposite to the conventional understanding that the flat band in cTBG has an identical physics with the LLL, where the presence of a FCI phase is expected. 

We further apply an analysis of the occupation number, like what we have discussed in the main text, there is an emergent Fermi liquid formed for $\nu\in (1/2, 2/3)$, while for $\nu\in (1/3, 1/2)$, the system shows a FCI behavior, even for $\nu=4/9$, which is quite close the phase transition at $\nu=1/2$. 
This thus aligns with our main conclusion.

\section{Details on the numerical simulations}
\label{sec:details_ed}

In this section, we provide the geometry we use for the rectangular and tilted sample. Note that we run all ED simulations in momentum space. 
Here, the rectangular sample means we use the moir\'e reciprocal vectors $\mathbf{G}_1$ and $\mathbf{G}_2$. 
For the tilted sample, the spanning vector is a linear combination of these two moir\'e reciprocal vectors \cite{PhysRevB.90.245401}.

All calculations on rectangular samples (Fig.~1, Fig.~2 in the main text, and Fig.~\ref{fig:other_fillings} in this supplemental materials), are with system size $(N_x, N_y)$ and spanning vectors $\mathbf{T}_1=(N_x,0)$ and $\mathbf{T}_2=(0, N_y)$.  

In the calculation of the charge density wave (Fig.~3 in this supplemental materials), we use two system sizes. 
For the case of $32$ sites, we use the spanning vectors $\mathbf{T}_1=(2,-4)$ and $\mathbf{T}_2=(6,4)$.
For the case of $36$ sites, we use the spanning vectors $\mathbf{T}_1=(6,0)$ and $\mathbf{T}_2=(0,6)$. 

In the calculation of the energy spectrum of tilted samples (Fig.~\ref{fig:tilted_result} in this supplemental materials), for band filling $\nu=3/7$ and $4/7$, we use a system size $28$ clusters, and the spanning vectors $\mathbf{T}_1=(4,-2)$ and $\mathbf{T}_2=(2,6)$. 
At band filling $\nu=4/9$ and $5/9$, we use a system size $27$ clusters, and the spanning vectors $\mathbf{T}_1=(6,3)$ and $\mathbf{T}_2=(3,6)$. 

We note that, the energy scale difference between the rectangular sample and the tilted sample, shown in Fig. \ref{fig:other_fillings} and Fig. \ref{fig:tilted_result}, is up to a constant in numerical simulations. 

\section{Hole dispersion and band geometry}
The Bloch wavefunction of an ideal Chern band takes the form $\psi_{\mathbf{k}}(\mathbf{r}) = N_{\mathbf{k}} B(\mathbf{r}) \phi^{\rm LLL}_{\mathbf{k}}(\mathbf{r})$ where $N_{\mathbf{k}}$ is a $k$-dependent normalization, and $B(\mathbf{r})$ is a $k$-independent quasi-periodic function with $\phi^{\rm LLL}_{\mathbf{k}}(\mathbf{r})$ the LLL wavefunction. Therefore the matrix elements $V_{\mathbf{k}_1\mathbf{k}_2\mathbf{k}_3\mathbf{k}_4}$ are related to the LLL form factors \cite{PhysRevLett.127.246403} as \begin{equation}
V_{\mathbf{k}_1\mathbf{k}_2\mathbf{k}_3\mathbf{k}_4} =   ( \prod_{i = 1}^{4} N_{\mathbf{k}_i} ) F_{\mathbf{k}_1\mathbf{k}_2\mathbf{k}_3\mathbf{k}_4}
\end{equation} where $F_{\mathbf{k}_1\mathbf{k}_2\mathbf{k}_3\mathbf{k}_4}$ is a function of the LLL form factors and the interactions. The normalization factor $N_{\mathbf{k}}$ is the major difference from a LL and controls the band geometry fluctuations. The Berry curvature $\Omega_{\mathbf{k}}$ and the FS metric is given by $|\Omega_{\mathbf{k}}| = |-1 + \Delta_{\mathbf{k}} \log  N_{\mathbf{k}}|={\rm tr} \>   g_{\mathbf{k}} $, where $\Delta_{\mathbf{k}}$ is the Laplacian operator. The hole energy is related to the normalization factor as \begin{equation}
\begin{aligned}
E_h(\mathbf{k}) = N_{\mathbf{k}}^2 \sum_{\mathbf{k'}} N_{\mathbf{k}'}^2 \big( F_{\mathbf{k} \mathbf{k}' \mathbf{k} \mathbf{k}'} + F_{\mathbf{k}' \mathbf{k} \mathbf{k}'  \mathbf{k}} - F_{\mathbf{k} \mathbf{k}' \mathbf{k}' \mathbf{k}} - F_{\mathbf{k}'  \mathbf{k} \mathbf{k} \mathbf{k}'}\big)  
\end{aligned}
\end{equation} 
So in ideal bands, there is an even closer connection between hole dispersion and the non-uniformity of the quantum geometry, directly from the normalization $N_{\mathbf{k}}$ which controls the fluctuations of Berry curvature and FS metric.

\section{Relation between Berry curvature of electrons and holes} 
In the main text, we have demonstrated the emergence of $\mathcal{C}_{\text{avg}}=1$ integer quantum anomalous Hall crystal after particle-hole transformation. In this section, we provide a detailed derivation of Berry curvature of electrons and holes in a given band.

Following the tenfold symmetry classification~\cite{PhysRevB.55.1142, RevModPhys.88.035005}, particle-hole symmetry relates the creation operator for an electron at momentum $\mathbf{k}$ and $m$th-band to the annihilation operator for a hole at momentum $-\mathbf{k}$ in the same band, $c^\dagger_{\mathbf{k},m}=d_{-\mathbf{k},m}$. In real space, this relation becomes $c^\dagger_{\mathbf{r},m}=d_{\mathbf{r},m}$. 
For simplicity, we now omit the band index $m$ in the following discussion.
The operators can be expanded in terms of Bloch wavefunctions as 
\begin{eqnarray}
    c^\dagger_{\mathbf{k}}=\int d\mathbf{r}e^{i\mathbf{k}\cdot\mathbf{r}}\mu^{e}_{\mathbf{k},\mathbf{r}}c^\dagger_{\mathbf{r}},~~d^\dagger_{\mathbf{k}}=\int d\mathbf{r}e^{i\mathbf{k}\cdot\mathbf{r}}\mu^{h}_{\mathbf{k},\mathbf{r}}d^\dagger_{\mathbf{r}},
\end{eqnarray}
we then have $\mu^{e}_{\mathbf{k},\mathbf{r}}=[\mu^{h}_{-\mathbf{k},\mathbf{r}}]^{*}$. From the viewpoint of Berry connection of holes $\mathbf{A}^{h}(\mathbf{k})=i\int d\mathbf{r}[\mu^{h}_{\mathbf{k},\mathbf{r}}]^{*}\nabla_{\mathbf{k}}\mu^{h}_{\mathbf{k},\mathbf{r}}$, this leads to
\begin{eqnarray}
    \mathbf{A}^{h}(\mathbf{k})=[\mathbf{A}^{h}(\mathbf{k})]^{*}=-i\int d\mathbf{r}[\mu^{e}_{-\mathbf{k},\mathbf{r}}]^{*}\nabla_{\mathbf{k}}\mu^{e}_{-\mathbf{k},\mathbf{r}}=\mathbf{A}^{e}(-\mathbf{k}).
\end{eqnarray}
In terms of Berry curvature, we then have
\begin{eqnarray}
    \Omega^{h}_{\mathbf{k}}=\nabla_{\mathbf{k}}\times\mathbf{A}^{h}(\mathbf{k})=-\nabla_{-\mathbf{k}}\times\mathbf{A}^{e}(-\mathbf{k})=-\Omega^{e}_{-\mathbf{k}}.
\end{eqnarray}
Together with the fact that creating a hole at momentum $\mathbf{k}^{h}$ equivalently means removing an electron at momentum $\mathbf{k}^{e}=-\mathbf{k}^{h}$, we have $\Omega^{h}_{\mathbf{k}^{h}}=-\Omega^{e}_{\mathbf{k}^{e}}$
In this sense, a fully occupied electron band has an opposite Chern number of a fully occupied hole band.

We now consider the case of a partially filled band. In a semiclassical picture, the Hall conductivity $\sigma_{xy}$ of electrons can be expressed as
\begin{eqnarray}
    \sigma_{xy}^{e}=-\frac{e^2}{h}\int_{\text{occ}} d\mathbf{k}\Omega^{e}_{\mathbf{k}}=-\frac{e^2}{h}(\mathcal{C}^{e}-\int_{\text{unocc}} d\mathbf{k}\Omega^{e}_{\mathbf{k}})=-\frac{e^2}{h}(\mathcal{C}^{e}+\int_{\text{occ. by holes}} d\mathbf{k}\Omega^{h}_{\mathbf{k}}).
\end{eqnarray}
where $\mathcal{C}^{e}$ is the Chern number of the fully filled band, which is $-1$ for the valence band of CTBG.
For the case discussed in the main text at electron band filling $\nu=1/4$, since the averaged many-body Chern number of ground states is trivial with ${C}_{\text{avg}}=0$ (and then $\sigma_{xy}^{e}=0$), the averaged many-body Chern number of ground states at hole band filling $\nu=3/4$ will satisfy $\mathcal{C}_{\text{avg}}=-\mathcal{C}^{e}=+1$, which is in line with the numerical calculations.

\end{document}